\documentclass[mathleft
]{an}

\usepackage{graphicx}
\usepackage{times}
\begin{document}

% The following seven commands are intended for editorial usage and should be ignored by
% the author(s).
\Pagespan{1}{ }% Document's page range. 
% If second parameter is left empty, the last page is computed automatically.
\Yearpublication{2009}%
\Yearsubmission{2009}%
\Month{}%   
\Volume{}%  
\Issue{}% 
 \DOI{}% 

\title{Populations of Supersoft X-ray Sources:\\
Novae,
tidal disruption,
Type Ia supernovae,
      accretion-induced collapse,
ionization, 
    and intermediate-mass black holes?        
}
%\thanks{Data
%from STELLA}}

\author{Rosanne Di\thinspace Stefano\inst{1}\fnmsep\thanks{Corresponding author:
  \email{rdistefano@cfa.harvard.edu}\newline}
%Example 
%for footnote, note the usage of the \texttt{fnmsep}
%command as separator between institute number and footnote mark} 
\and  Francis A. Primini\inst{1}
\and  Jifeng Liu\inst{1}
\and  Albert Kong\inst{2}
\and  Brandon Patel\inst{1,3}
}
\titlerunning{Populations of Supersoft X-ray Sources}
\authorrunning{Di\thinspace Stefano et al.}
\institute{Harvard-Smithsonian Center for Astrophysics 
\and 
National Tsing Hua University
\and 
Rutgers University}

\received{30 July 2009}
%\accepted{later}
%\publonline{later}

\keywords{
X-rays: binaries -- white dwarfs -- nova, cataclysmic variables -- 
stars: neutron -- black hole physics -- X-rays: ISM -- galaxies: general -- 
supernovae: general   
}

\abstract{%
Observations of hundreds  of supersoft  x-ray sources (SSSs) in external 
galaxies have shed light on the diversity of the class and on the
natures of the sources. SSSs are linked to the physics of Type~Ia 
supernovae and accretion-induced collapse, ultraluminous x-ray sources
and black holes, the ionization of the interstellar medium,
and tidal disruption by supermassive black holes. 
The class of SSSs has an extension to higher luminosities: ultraluminous
SSSs have luminosities above $10^{39}$~erg~s$^{-1}$. There is also
an extension to higher energies: quasisoft x-ray sources (QSSs) emit
photons with energies above $1$~keV, but few or 
none with energies above $2$~keV.
Finally, a significant fraction of the SSSs found in external galaxies
switch states between observations, becoming either quasisoft or hard. 
For many systems ``supersoft'' refers to a temporary state; SSSs
are sources, possibly including a variety of fundamentally different
system types, that pass through such a state.  
We review 
those results derived from extragalactic data and related theoretical work 
that are most surprising and that suggest directions
for future research.   
}

\maketitle

\section{Introduction} 

\subsection{Nearby SSSs: A small, exclusive group}  

Supersoft x-ray sources (SSSs) are unusual among x-ray sources
because they emit few photons with energies above $1$~keV.
They are unusual in another way: the definition of the class
is very general and can in principle encompass many different types of
physical systems.

SSSs were established as a class by {\it ROSAT} in the early 1990s. 
The original definition of SSSs is often expressed as follows:
luminosities in the range of approximately
 $10^{36}-10^{38}$~erg~s$^{-1}$ and effective
temperatures with $k\, T$ in the range of tens of eV.
This phenomenological definition is based on the detection of 
a small number of sources. By 1996,  
seven SSSs had been discovered in the Galaxy, and eleven in the 
Magellanic Clouds (Greiner 1996).

The empirical definition of SSSs suggests effective radii
comparable to the radii of white dwarfs. 
Indeed, some SSSs are associated with hot white dwarfs in recent
novae or symbiotic binaries, or with the central stars of planetary nebulae.

Not all SSSs are associated with
systems known to contain white dwarfs. 
Roughly half of those with optical IDs
are binaries with orbital periods ranging from hours to a few days.
These close-binary supersoft sources (CBSSs) may be white dwarfs accreting at 
high enough rates to allow incoming matter to undergo quasisteady nuclear
burning. van den Heuvel et al.\, (1992) developed a model in which
a Roche-lobe filling donor which is more massive than the white dwarf and/or
slightly evolved can provide mass
at the requisite high rates ($\sim 10^{-7} M_\odot$~yr$^{-1}$).   

The CBSS model was important not only as
a possible explanation of SSSs, but also because it provided a channel
through which white dwarfs could retain accreted matter, thereby
avoiding nova explosions that would deplete the white dwarf.
Genuine mass retention by a white dwarf is exciting because it provides a 
way for some white dwarfs to achieve the Chandrasekhar mass
and to undergo accretion-induced collapse (AIC; van den Heuvel et al.\, 1992)
or Type~Ia supernova explosions (Rappaport, Di\thinspace Stefano \& Smith
1994).  

Although the white dwarf models seemed ideally suited to the first group of
 SSSs that
had been discovered, other explanations were naturally explored as well.  
Kylafis and Xilouris (1993) considered a neutron star model in which the
photospheric radius is comparable to the radius of a white dwarf.

Stellar-mass black holes sometimes exhibit states
referred to as ``high soft'' or ``thermal dominant'' states 
(Remillard \& McClintock 2006).
In these cases, however, the effective values of $k\, T$ are around
$1$~keV, making the sources much harder than SSSs. Nevertheless,
CAL~87, a Magellanic Cloud source often referred to as one of the
``classical SSSs'', 
has been considered as a black hole candidate 
(Cowley et al.\, 1990).

During the past $17$~years, more information has been gathered
about the SSSs in the Galaxy and in the Magellanic Clouds. 
Much of the new data is consistent with white dwarf models.
Yet, definite identification of the nature of the accretors
has proved difficult.

 \subsection{External Galaxies: An extended arena} 
 
The small population of 16 SSSs found by {\it ROSAT} in
M31 represents a total population that could be as large
 as $\sim 1000$ (Di\thinspace Stefano \& Rappaport 1994). 
The advent of {\it Chandra} and {\it XMM-Newton} 
 provided the opportunity to discover more of the M31 SSSs
and to identify SSSs in a variety of other external galaxies. 
  Discovering more SSSs allows us to establish the full range
of source properties and the distributions of temperatures  and 
luminosities. Clues 
to the natures of the sources can be gathered by studying their location
 within galaxies. For example, are SSSs associated primarily with old
or young stellar populations?Are the numbers of SSSs in
typical galaxies large enough to support the hypothesis that they
 are the dominant class of Type~Ia progenitors?

After a decade of extragalactic x-ray surveys, we can begin
 to answer some of these questions. Beyond this, we have made discoveries along new lines that
reveal SSSs to be an even more interesting class of
sources  than many of us anticipated. In this short paper
we summarize the results that are most surprising and that
suggest new lines of investigation.
   
\section{Ten Years of Surprises}
\subsection{SSSs Near Galactic Nuclei}  

Only two external galaxies in the Local Group, M31 and M32,
are known to have supermassive black holes.
Because of the relative proximity of these galaxies, {\it Chandra}
is able
to resolve x-ray sources with $L_x$ greater than approximately\\   
$10^{36}$~erg~s$^{-1}$ within a few arcseconds of the nucleus.  
In M32, Ho et al. (2003) find three sources
within $30^{\prime\prime}$ of the galactic center. One of these is   
the nucleus itself, a $2.5 \pm 0.5 \times 10^6 M_\odot$ black hole.
The source closest to the nucleus 
 lies $1.5^{\prime\prime}$ away, roughly $7$~pc, in projection. 
It is an SSS which can be fit by a $40$~eV thermal model  
or by a power-law model with $\Gamma=9.2^{+1.7}_{-1.4}$ and
$L_X=2.2\times 10^{36}$~erg~s$^{-1}$. 
Because SSSs constitute a relatively small fraction of all x-ray sources,
it seems remarkable that the closest x-ray source to the supermassive black
hole happens to be an SSS. Interestingly enough, however, the situation
in M31 is similar. Within a few arcseconds of
the black hole is a soft source,
marginally hotter than the ``classical'' SSSs known in the Magellanic
Clouds (Di\thinspace Stefano rt al.\, 2004).
It is unlikely to be coincidental that 
very soft sources lie so close to the central black hole in the only
two galaxies within which such a phenomenon can be detected. 
There may be
several possible explanations. 

One possibility is that the soft sources are the cores  
 of giant stars that have been tidally disrupted by the supermassive black hole 
(Di\thinspace Stefano et al.\, 2001).
Tidal disruption is a well-studied 
phenomenon from the theoretical
perspective (Rees 1988; Loeb \& Ulmer 1997; Ulmer 1999). 
The flare which follows such a disruption
can be bright ($> 10^{44}$~erg~s$^{-1}$), and can last for days to months.
In recent years, flare events that may be associated with tidal 
disruptions in distant galaxies have been detected (see, e.g., Esquej 
et al.\, 2008).
When the disrupted star is a giant\footnote{Only giants can be
disrupted by black holes with mass larger than about $10^8 M_\odot,$
because main sequence 
stars cross the event horizon before being disrupted.},
the remnant is a long-lasting ($> 10^3$~year) hot core
that could be detected as an SSS. The computed
rate of disruptions (Magorrian \& Tremaine 1999)
 is high enough that large galaxies may
each house several stripped cores. 

Whatever the natures of the soft sources in the nuclei of
M31 and M32, the tidal stripping of cores is a definite prediction,
and must produce an excess of soft x-ray sources in the central
regions of some galaxies. The sources may be on
orbits that bring them far from the nucleus, even
while still
hot enough to be detected at x-ray wavelengths  
This means that 
tidal disruption can have long-lasting
consequences that are detectable in nearby galaxies, even though
flares are rare and are therefore preferentially observed in very 
distant galaxies.

This example shows that some SSSs in galaxy centers may have 
evolutionary paths different
from those computed for supersoft sources descended from primordial binaries. 
 
\subsection{SSSs in Young Stellar Populations}  

In the CBSS model derived by van den Heuvel et al.\, (1992), the requisite
high rate of accretion is produced by the thermal time scale
adjustment of the donor star to a shrinking Robe lobe. Dynamical
stability dictates that the initial donor mass be less than roughly 
$2.5\, M_\odot$ (see, e.g., Langer et al.\, 2000).  
In a second class of CBSS models, the white dwarf
is fueled by irradiation-driven winds 
from a donor that is even less massive  
(van Teeseling \& King 1998).  
Given this mass range, 
most CBSSs should begin their
period of high mass transfer and possible SSS-like activity
more than $10^9$~years after the binary is formed. The primary must 
first have a chance to evolve to become a white dwarf. Once it does,
there is generally a wait time before the donor can begin to contribute
mass. The orbit must shrink (most likely to occur through
magnetic braking) and/or the donor must begin to evolve. In either case, the
binary has time to move away from the place where it formed.
We would therefore not expect SSSs to be primarily associated with
regions containing young stars. 

\begin{figure}
\includegraphics[width=80mm,height=80mm]{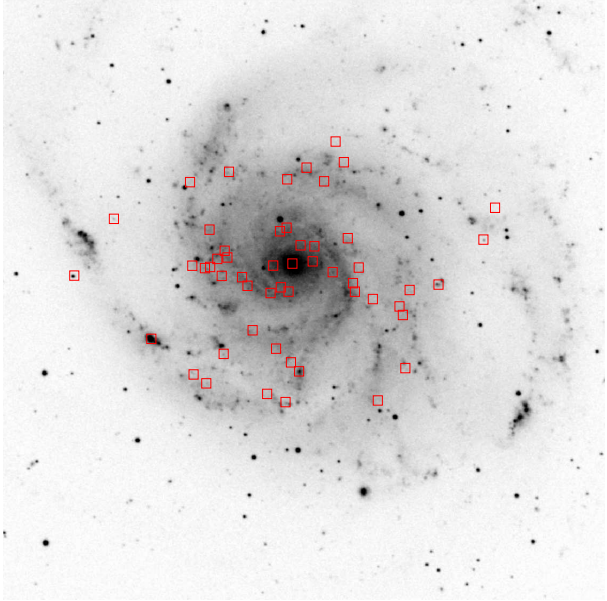}
\caption{SSSs in the spiral galaxy M101. 
SSSs are found in the bulge, but  most cluster in or near
the spiral arms. Whatever their nature(s), SSSs can ionize
the ISM (Rappaport et al.\, 1994; Chiang \& Rappaport 1996).
So far, searches for nebulae associated with SSSs have had limited
success (Pakull \& Motch 1989; Remillard et al.\, 1995)}
\label{label1}
\end{figure}
\begin{figure}
\includegraphics[width=80mm,height=80mm]{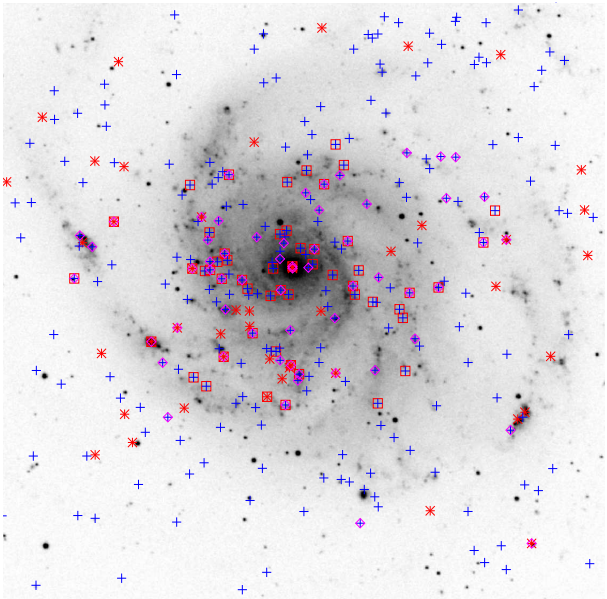}
\caption{All x-ray sources in M101. The position of each detected source is marked with a blue cross. As above, red squares mark the positions 
of SSSs. Red stars correspond to x-ray sources with 
$L_X > 10^{38}$~erg~s$^{-1}.$  
Red diamonds are for quasisoft sources (\S 2.3.2 and Figure 3) {\bf Note that
the distribution of x-ray sources is far more extended than the
distribution of SSSs. Furthermore,  many x-ray sources (and few SSSs)
occupy the area between spiral arms.}}
\label{label1}
\end{figure}

It was therefore surprising to find, even in the first 
{\it Chandra} observations of the disk of M31
(Di\thinspace Stefano et al.\, 2004), that 
SSSs in the disk are clustered near star-forming regions,
possibly indicating that they are young. 
This result was confirmed and strengthened
by observations of other galaxies (Di\thinspace Stefano \& Kong 2004).
Liu (2008) has identified 
SSSs in 383 galaxies. The location of SSSs in M101 relative
to other  sources and to the optical features of the galaxy
is shown in Figures 1 and 2. 
 
The connection with star-forming regions 
confirms that many sources are 
young. 
The donors are likely to be significantly more
massive than white dwarfs. They therefore cannot stably 
transfer mass to a white dwarf through the L1 point.
Instead, they are
likely to be contributing mass through a wind.
Many SSSs near star-forming regions may be high-mass x-ray binaries
(HMXBs) or symbiotic binaries.

\subsection{Empirical Extensions of the Class} 

One of the surprises of population studies is that the class of 
SSSs has extensions to both higher luminosities
($> 10^{39}$~erg~s$^{-1}$) and 
higher energies ($k\, T$ in the range of a $150-350$ eV).

\subsubsection{Ultraluminous SSSs (ULSs)}

Sources with
x-ray luminosities above  $10^{39}$~erg~s~$^{-1}$ are referred to as   
ultraluminous x-ray sources (ULXs).
Because their luminosities are larger than the Eddington luminosity for
a ten-solar-mass object, it has been suggested that some may
be intermediate-mass black holes (IMBHs). 
Some SSSs are ultraluminous; such sources are sometimes referred
to as ULSs.
Several of these sources are well studied. (See, e.g.,
Fabbiano et al.\, 2003;
 Soria \& Ghosh 2009; 
Mukai et al.\, 2005;  Kong et al.\, 2004;
Kong \& Di\thinspace Stefano 2005). 
It has been suggested that some are white dwarfs
in super-Eddington outbursts and/or white dwarfs with beamed emission.
Stellar-mass and intermediate-mass black holes have also been
considered.

\subsubsection{Quasisoft x-ray sources (QSSs)} 
When we started to look for SSSs in external galaxies, we expected 
that there would be a gap 
between their spectra and the spectra of the canonical hard sources
normally associated with neutron star and black hole accretors.
Instead, we found a continuum of soft source energies.

In fact, many sources 
provide some photons above
$1$~keV, while exhibiting no emission above $2$ keV.  
Sources with $k\, T$ near $100$~eV can have such spectral properties,
particularly if they are highly absorbed. Such sources are   
important, even if one is interested 
only in white dwarfs, because they could be
accreting white dwarfs with mass close to the 
Chandrasekhar mass. 

It could therefore have been the case 
that the harder soft sources were merely
highly absorbed hot white dwarfs. 
To test this hypothesis, we identified all
of the sources in several external galaxies, with the goal of finding
a significant number of soft sources bright enough for reliable spectral fits. 
This process identified both absorbed sources with $k\, T$ in the 
supersoft range and sources that are intrinsically hotter. 
(Di\thinspace Stefano et al.\, 2006)  
Indeed, some sources 
have fits with $k\, T \sim 250-350$~eV. 
These sources are not natural candidates for nuclear-burning 
white dwarfs (NBWDs). 
Sources with luminosities above $10^{36}$~erg~s$^{-1}$ which  produce 
few or no photons with
energies above $2$~keV are called quasisoft x-ray sources (QSSs).  
QSSs represent a high-energy extension of the class of SSSs.
Like SSSs, QSSs have been discovered in every galactic environment.  
(See Figure 3 and Table 1.) 
\begin{figure}
\includegraphics[width=80mm,height=80mm]{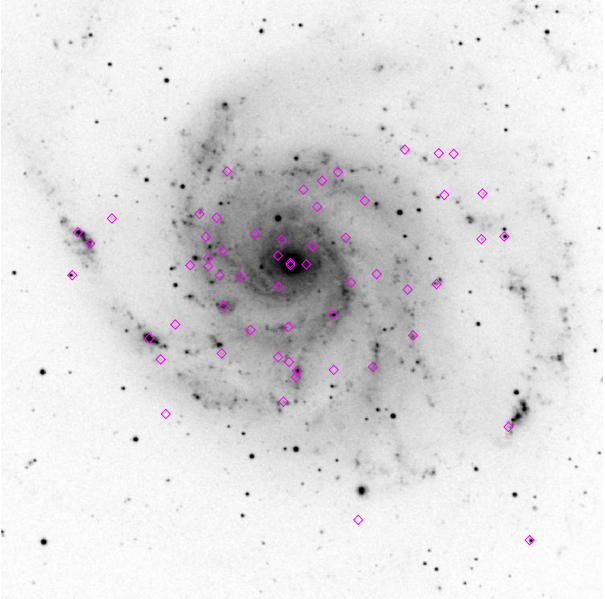}
\caption{QSSs in the spiral galaxy M101. Note that QSSs also appear to
be concentrated  near the spiral arms, although they are somewhat
more spread out than SSSs.  
}
\label{label1}
\end{figure}

\subsubsection{State-Changing SSSs} 

The eighteen SSSs comprising the first-discovered 
set have been observed at occasional intervals over a period of
$\sim 20$ years. Although changes in flux
are common, none of the ``classical'' SSSs  
have been reported maintain a high luminosity while switching to 
a harder spectal state.
Among the much larger number of sources observed 
in external galaxies, however, several state changers have been
well-documented. One of these is M101-ULS-1. 
As Figure 4 shows, this source has been detected in supersoft,
quasisoft, and hard states. It is very likely a black hole,
either of stellar or intermediate mass (Mukai et al.\, 2005;  
Kong et al.\, 2004;
Kong \& Di\thinspace Stefano 2005). 
Two state-changers have been discovered in M31
(Pietsch et al.\, 2005; Orio 2006; Patel et al.\, 2009). These
are different from M101-ULS-1 in that they are sources of much
lower luminosity. They are not likely to be black holes.
 The light curve of one of the M31 sources is
shown in Figure 5.      
\begin{figure}
\includegraphics[width=80mm,height=80mm]{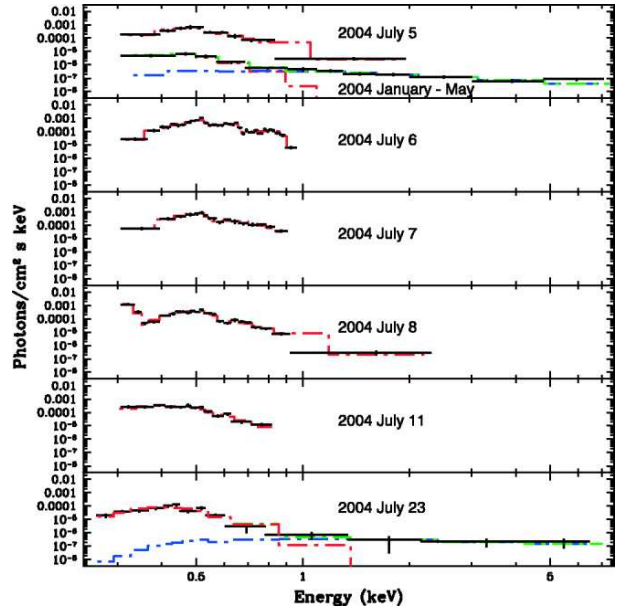}
\caption{
Energy distribution of photons received from M101-ULS-1.
Note that during some observations all of the photons have energies
below $\sim 1$~keV; the source is supersoft. In others, there is emission
above $1$~keV, but none above $2$~keV; the sources is quasisoft. The source
also passes through hard states with lower luminosity than during the 
soft states. M101-ULS-1 is a considered to be a black hole candidate. 
Stellar-mass and 
intermediate-mass models have been considered.  
}
\label{label1}
\end{figure}
\begin{figure}
\includegraphics[width=80mm,height=80mm]{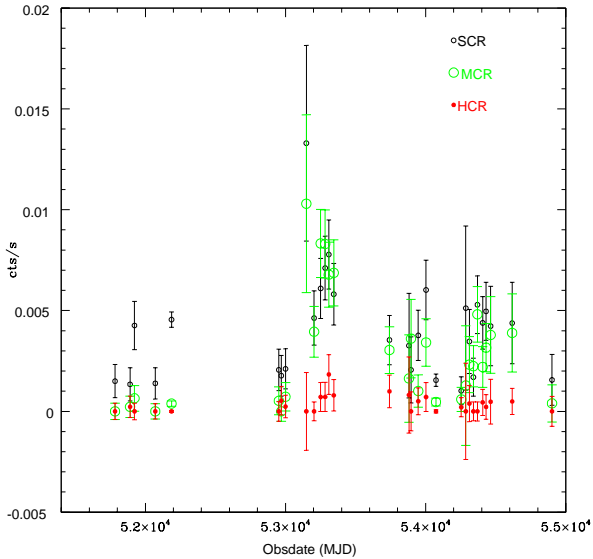}
\caption{The light curve of r1-25, an x-ray source in M31.
{\bf Black:} count rate in the soft band (0.1-1.1 keV);
{\bf Green:} count rate in the medium band (1.1-2 keV);
{\bf Red:} count rate in the hard band (2-7 keV). Note that
the soft band dominates during the first set of observations, with few
or no counts in the medium and hard bands. The source is SSS. In later
observations, the count rate is higher in the medium band; during some
observations 
there are significant detections in the hard band.
This source passes through SSS, QSS, and hard
states. The estimated luminosity in all cases is less than
$10^{38}$~erg~s$^{-1}$.  
}
\label{label1}
\end{figure}

\subsection{Size of the Population}
\begin{table}
% \centering%%%
\caption{Soft Sources in External Galaxies}
\label{tlab}
\begin{tabular}{cccc}\hline
Galaxy  & SSSs & QSSs & Other Sources \\
\hline 
M101       & 42  & 21  & 24   \\
M83        & 28  & 26  & 74   \\
M51        & 15  & 21  & 56   \\
M104       & 5   & 17  & 100  \\
NGC4472    & 5   & 22  & 184  \\
NGC4697    & 4   & 15  & 72   \\
\hline
\end{tabular}
\end{table}

SSSs and QSSs have now been discovered in every galactic environment:
early type galaxies, in both the bulges and spiral arms of late-type galaxies,
and in globular clusters. Table 1 shows the results for a set of six
galaxies that have each been carefully analyzed.  
M101, M83, and M51 are spiral galaxies. M104 is a bulge-dominated spiral,
while both NGC4472 and NGC4697 are elliptical galaxies. Very soft
sources (either SSS or QSS) constitute a larger fraction of all x-ray
sources in late-type galaxies. For SSSs there is a dramatic decline in their
numbers for early-type galaxies relative to late-type galaxies. 

An automated source selection and identification process was employed
by Liu (2008) to study x-ray sources in fields containing
383 nearby galaxies. SSSs and QSSs were identified using the
same algorithm used for the galaxies in Table 1. 
Liu found that $2.6\%$ of all sources 
bright enough for spectral classification
are SSS. For every SSS there are four QSSs. The combination of SSSs and
QSSs constitute about $13\%$ of all x-ray sources.
The high ratio of QSSs to SSSs likely reflects the prevalence of
older stellar populations among the galaxies in Liu's survey.
In addition, Liu estimates that $\sim 22\%$ of the SSSs are state-changers.

Among the SSSs alone, Liu identifies $5$ sources with
$L_X > 5 \times 10^{39}$~erg~s$^{-1}$, and $22$ sources with
$L_X > 1 \times 10^{39}$~erg~s$^{-1}$. 
In estimating the luminosities, Liu has used a power-law model,
which may underestimate the luminosity of the SSSs. 

\subsection{SSSs as Progenitors of Type Ia Supernovae}

In order for an accreting white dwarf to achieve the Chandrasekhar mass
and explode as a Type~Ia supernova, it must generally gain at
least $0.2\, M_\odot.$ Retention of matter by WDs appears to
require nuclear-burning.  
The burning of matter releases a great deal of energy, 
$> 10^{38}$~erg~s$^{-1}$ for white dwarf masses near $M_C.$ 
In the simplest model, these sources should radiate as SSSs, with 
$k\, T> 80$~eV. 

The progenitors of Type~Ia supernovae should be the hottest and brightest  
SSSs. 
Di\thinspace Stefano \& Rappaport (1994) showed that those SSSs which
are Type~Ia progenitors could be detected in nearby galaxies with high 
efficiency. Because it takes time ($> 2\times 10^5$~yr) to accrete   
$> 0.2\, M_\odot$ at rates of $\sim 10^{-6} M_\odot$~yr$^{-1},$  
the numbers of actively accreting, hot, bright progenitors in
galaxies such as our own and M31 must be large.
\begin{equation}
N = 
750\,   
  \Big({{\Delta\, M}\over{0.2\, M_\odot}}\Big)  
\Big({{8\times 10^{-7} \frac{M_\odot}{yr}}\over{\beta \, \dot M_{in}}}\Big)  
\Big({{L_B}\over{10^{10} L_\odot}}\Big).
\end{equation}
For most spiral galaxies
(and even for ellipticals, with a lower rate of SNe Ia),
$N$  cannot be much smaller than several hundred, and is likely to
be larger.

Comparing the number of hot, bright NBWDs
needed to account for the expected Type~Ia supernovae to the numbers
of SSSs and QSSs actually detected in each galaxy, we find a discrepancy
larger than a factor of ten. In fact, the true discrepancy is almost
certainly larger. This is because not all SSSs, and possibly only a small
fraction of QSSs, are NBWDs. Furthermore, of those soft sources that are
NBWDs, many are too dim and cool for the white dwarf to have a mass near
$M_C.$  

Thus, x-ray observations of external galaxies falsify the hypothesis that
the majority of Type~Ia supernovae are generated by NBWDs that are
detectable as SSSs during the epoch in which the white dwarf's mass is 
increasing.  There are two possibilities: either
the majority of Type Ia progenitors are not NBWDs approaching
the Chanadrasekhar limit, or
they are, but we cannot detect them as SSSs.

In fact, it may be the case that those accreting white dwarfs 
on the way to becoming Type~Ia supernovae tend to eject more
winds, which can then block the soft radiation and obscure the source.
In addition, a low duty cycle of activity, expected for recurrent novae,
could also make the sources less likely to be detected in a supersoft phase.

\subsection{SSSs and Accretion-Induced Collapse} 

The most serious bottleneck found in 
binary evolution calculations of accreting white dwarfs
that might be Type Ia progenitors 
is the difficulty in channeling enough mass to the accretor to
allow it to achieve the Chandrasekhar mass. This problem is 
less severe for those white dwarfs that start
with masses within a few tenths of $M_C.$ These tend to be O-Ne
white dwarfs which experience AIC when reaching the Chandrasekhar
mass, instead of Type Ia explosions.

Because the white-dwarf progenitors of AIC must be massive,
the binaries in which they form must be wide. Furthermore, the
companion which donates mass to the white dwarf is likely to be massive.
AICs are therefore likely to take place near star-forming regions.
Before the collapse, the white dwarf is likely to be accreting
winds ejected by the donor star, with matter infalling at rates
of roughly $10^{-6} M_\odot$~yr$^{-1}$. The white dwarf
should appear as an SSS,
at least episodically. After the collapse, the winds will continue.
The newborn accreting neutron star  
will be highly luminous and may appear as a ULX.
This model (Di~Stefano, Pfahl, \& Harris 2009) links the SSSs found near
star forming regions with a subset of ULXs. 
ULXs are also preferentially found  
near star-forming regions. 

\subsection{Extensions of the Models} 

At the time SSSs were discovered, NBWDs 
provided the most natural explanation for their observed ranges of 
luminosities and temperatures.
Indeed, the significant association between novae and SSSs
verifies that many transient SSSs are indeed hot white dwarfs
(Pietsch et al.\, 2005), while
observations of nearby CBSSs 
provide evidence that NBWD 
models are likely to apply. 

In the time since the early 1990s, however, two developments argue
for extensions of the physical models. The first development is the
 extension of the ranges of
source temperatures and luminosities described in \S 2.4. 
The discovery of QSSs and ULSs provide examples of systems that are 
too hot and/or too luminous to be white dwarfs. 

The second development is the consideration of higher-mass black holes.
First, black holes with masses between $15\, M_\odot$ and $33\, M_\odot$
(Orosz et al.\, 2007; Prestwich et al.\, 2007) have been discovered.
Second,  
IMBH models have been developed as explanations
of ULXs. For black hole accretors with masses larger than a 
few solar masses, an optically thick inner disk can emit as a QSS or SSS.
(See Figure 6.) If the system is in a thermal dominant state, the state
in which it is possible determine the black hole mass,  it
may appear as a   
QSS or, for more massive black holes, as an SSS.   
(See Di Stefano et al.\, 2003a, 2003b, 2004a, 2004b; 
Di\thinspace Stefano \& Kong 2003a, 2003b, 2004a, 2004b)

While back hole models are natural for some QSSs and SSSs, neutron star
models should also be reconsidered. During the early 1990s, when
the only known neutron
star accretors were hard x-ray sources, it seemed to require fine tuning
to achieve a white-dwarf-like photospheric radius. The discovery of QSSs
allows a wider range of photospheric radii, perhaps removing 
the need for fine tuning.
In addition, white dwarf models in which a cooling white dwarf produces
soft radiation, but harder emission is produced through accretion, can also
be used to model state-changing soft sources (Patel et al.\, 2009).

\smallskip

\noindent{\bf Conclusions:} SSSs 
are found in all galactic environments. 
Many are young stellar systems. 
Several evolutionary paths can produce soft sources, while not all NBWDs,
even those of high mass, are detectable as SSSs.
The sum of these discoveries suggest that soft x-ray sources provide a rich area
for research on many astrophysical topics 
of great interest.  

\smallskip

\noindent{\bf Acknowledgments:} This work was supported 
in part by NASA through 
AR-10948.01-A-R and GO8-9092X.  

\begin{figure}
\includegraphics[width=80mm,height=120mm]{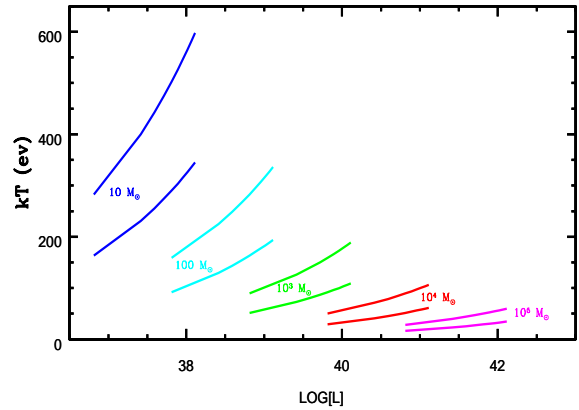}
\vspace{-2.3 true in}
\caption{$k\, T$ versus $Log[L]$ for the inner portion of the accretion disk
around black holes. Each pair of two curves of a single color corresponds
to a fixed black hole mass which labels the regions between the curves.
The upper curve of each color corresponds to a disk  with inner radius $6 M_{BH} G/c^2,$
while the lower curve 
corresponds to an inner disk with $3$ times the radius.
 The point at the bottom (top) of each curve corresponds to
the luminosity of the inner disk being $1\% L_{Eddington}$
($10\% L_{Eddington}$).    
}
\label{label1}
\end{figure}

\vspace{-.3 true cm}

\end{document}